\documentclass[aps,prl,reprint,groupedaddress]{revtex4-1}

\usepackage{graphics}
\usepackage{graphicx}
\usepackage{epsfig}

\begin{document}

\title{Phase Space Factors for Double Beta Decay: an up-date}
 
\author{Mihail Mirea$^{a,b}$, Tudor Pahomi$^{a,c}$, Sabin Stoica$^{a,b}$}

\affiliation{a) Horia Hulubei Foundation, P.O. MG12, 077125-Magurele, Romania\\
b) Horia Hulubei National Institute of Physics and Nuclear Engineering, P.O. Box MG6, 077125-Magurele, Romania\\
c) University of Bucharest, Faculty of Physics, P.O. Box MG11, 077125-Magurele, Romania}

\begin{abstract}
\noindent
 We give a complete, up-date list of the phase space factors (PSF) for $\beta^-\beta^-$, $\beta^+\beta^+$, 
$EC\beta^+$ and $ECEC$ double beta decay (DBD) modes, in all nuclei of interest and possible transitions to final states. In calculation, the Coulomb distortion of the electron wave functions is treated by solving numerically the Dirac equation with inclusion of the finite nuclear size and electron screening effects. In addition to the previous calculations we use a Coulomb potential derived from a realistic proton density distribution in nucleus, improve the precision of the numerical routines used to solve the Dirac equations and to integrate the PSF expressions, and use recently reported $Q$-values. These ingredients proved to be important, leading in many cases to significant differences as compared to the present available PSF values, which are discussed as well. 
Accurate values of the PSF are necessary ingredients both for theorists, to improve the DBD lifetime predictions and constraint the neutrino parameters, and for experimentalists to plan their set-ups. 

\end{abstract}


\maketitle

\section{Introduction}
\label{1}

Double beta decay is the most rare nuclear process measured so far which presents a great interest, especially for testing the lepton number conservation (LNC) and understanding the neutrino properties. Within the Standard Model (SM) it conserves the lepton number and can occur through several decay modes, with emission of two neutrinos/anti-neutrinos ($2\nu$) in the final states. However, theories beyond SM predict that this process may also occur without conservation of the lepton number, and hence, without emission of neutrinos/anti-neutrinos, through the so called neutrinoless ($0\nu$) decay modes.  According with the number and type of leptons we may have the following DBD modes: i) two neutrino double-electron decay ($2\nu\beta^-\beta^-$); ii) neutrinoless double-electron decay ($0\nu\beta^-\beta^-$); iii) two neutrino double-positron decay ($2\nu\beta^+\beta^+$); iv) neutrinoless double-positron decay ($0\nu\beta^+\beta^+$); v) two neutrino electron capture positron emitting decay ($2\nu EC\beta^+$);  vi) neutrinoless electron capture positron emitting decay ($0\nu EC\beta^+$); vii) two neutrino double electron capture decay ($2\nu ECEC$) and viii) neutrinoless double electron capture decay ($0\nu ECEC$).
 Complete information about the achievements in the study of DBD can be found in several recent excellent reviews \cite{AEE08}-
\cite{VES12}, which also contain a comprehensive list of references in domain. The lifetimes for the DBD modes can be factorized, in good a approximation, as follows: 

\begin{eqnarray}
 \left( T^{2\nu}_{1/2} \right)^{-1}=G^{2\nu}(E_0, Z)\mid M^{2\nu}\mid^2~~;~~
\left( T^{0\nu}_{1/2} \right)^{-1} \nonumber\\
=G^{0\nu}(E_0, Z)\mid M^{0\nu}\mid^2 \left(< \eta_{l}> \right)^2 \ ,
\end{eqnarray}

\noindent
where $G^{(2\nu, 0\nu)}$ are the PSF and $M^{(2\nu, 0\nu)}$ the nuclear matrix elements (NME) for the $2\nu$, $0\nu$ decay modes respectively, and $<\eta_l>$ is a beyond SM parameter which contains information about the neutrino properties in the (most common) hypothesis that the $0\nu$ $\beta\beta$ decay (DBD) occurs through the neutrino exchange mechanism. As seen, PSF and NME are key quantities for estimating/predicting DBD lifetimes and/or for deriving neutrino properties, hence it is important to calculate them with high accuracy. Moreover, the PSF values largely fix the order of magnitude of the DBD lifetimes. So far, much effort has been and is still devoted to the precise calculation of the NME \cite{ROD07}-\cite{RAH10}, while PSF have been less discussed because it was considered that they were computed with enough precision \cite{PR59}-\cite{SC98}.  Recently, the PSF were recalculated with a more rigorous method, by using exact electron Dirac wave functions (w.f.) and taking into account the finite nuclear size and electron screening effects \cite{KI12}-\cite{SM13}. The authors  found differences between their results and those calculated previously with approximate electron w.f., especially for heavier nuclei and for $\beta^+\beta^+$ decay modes. This justifies a more careful independent re-evaluation of the PSF involved in DBD, using more accurate methods. 

The purpose of this work is to give a complete, up-date list of the PSF values for the i)-vii) DBD modes mentioned above, in all nuclei of interest and possible transitions to final states. The decay mode $0\nu ECEC$ can not occur to the order of approximation that is presently considered in the literature. We developed high precision routines to compute the relativistic (Dirac) electron w.f. taking into account  the nuclear finite size and screening effects. In addition to the previous calculations, we use a Coulomb potential derived from a realistic proton density distribution in nucleus, improve the precision of the numerical routines used to solve the Dirac equations and integrate the PSF expressions, and use $Q$-values reported recently \cite{audi12}. These ingredients proved to be important, leading in many cases to significant corrections to the present available PSF values, which are discussed as well. Accurate PSF values (besides the NME) are necessary ingredients both for theorists to improve the DBD lifetime predictions and constraint the neutrino parameters, and for experimentalists to plan their set-ups. 

\section{Theoretical framework}

To compute PSF for DBD decay modes we need first to obtain the w.f. of the electron(s)/positron(s) emitted or electron(s) captured in the decay, which are distorted by the Coulomb potential of the nucleus. Older calculations have used a non-relativistic approach where the distortion of the w.f. by the Coulomb field was considered through Fermi (Coulomb) factors obtained (for the emitted particles) by taking the square of the ratio of the Schr\"odinger scattering solution for a point charge $Z$ to a plane wave, evaluated at the origin \cite{PR59}-\cite{HS84}. In a better approximation, the Fermi factors are calculated using a relativistic treatment of the electron/positron w.f, but with approximate Dirac functions (the Fermi factor is defined as the square of the ratio of the values of the Dirac w.f. of the electron at the nuclear surface) and without the inclusion of screening effects \cite{DOI83,SC98}. Recently, Kotila and Iachello (KI) recalculated the PSF using exact Dirac electron/positron w.f. and including the screening effect \cite{KI12}-\cite{KI13}. In this work we adopt this more rigorous relativistic treatment of KI but with the inclusion of additional ingredients, as it is described in the following.

\subsection{The radial wave functions}

For free states we use relativistic scattering electron/positron w.f. solutions of the Dirac equation in a central (Coulomb) potential:

\begin{equation}
\Psi^+_{\epsilon\kappa\mu}(r) = \left(\begin{array}{l}g_{\kappa}(\epsilon,r)\chi^\mu_\kappa\\if_\kappa(\epsilon,r)\chi^\mu_{-\kappa}\end{array}\right)~~ 
\end{equation}
for $~\beta^- ~\rm{decay}$ and 
\begin{equation}
\Psi^-_{\epsilon\kappa\mu} =\left(\begin{array}{l}if_{\kappa}(\epsilon,r)\chi^{-\mu}_{-\kappa}\\-g_\kappa(\epsilon,r)\chi^{-\mu}_\kappa\end{array}\right)~~i\end{equation}
 for $\beta^+~ \rm{decay}$
\noindent
where $\kappa=(l-j)(2j+1)$ is the relativistic quantum number and $\chi^\mu_\kappa$ are spherical spinors. The quantities 
$g_{\kappa}(\epsilon,r)$ and $f_{\kappa}(\epsilon,r)$ are the large and small components of the radial w.f. which satisfy the radial equations:

\begin{eqnarray}
\label{dirac}
{dg_{\kappa}(\epsilon,r)\over dr}=
-{\kappa\over r}g_{\kappa}(\epsilon,r)
+{\epsilon-V+m_ec^2\over c\hbar}f_{\kappa}(\epsilon,r)\\
{df_{\kappa}(\epsilon,r)\over dr}=
-{\epsilon-V-m_ec^2\over c\hbar}g_{\kappa}(\epsilon,r)
+{\kappa\over r}f_{\kappa}(\epsilon,r) \nonumber
\end{eqnarray}
where $V$ can be negative/positive (for $\beta^-/\beta^+$).
 These functions are normalized so that they have 
the following asymptotic behavior:
\begin{eqnarray}
\label{asi}
\left(\begin{array}{l}g_{k}(\epsilon,r)\\f_{k}(\epsilon,r)\end{array}\right)\sim
{\hbar e^{-i\delta_k}\over pr}
\left(\begin{array}{l}\sqrt{{\epsilon+m_ec^2\over 2\epsilon}}
\sin(kr-l{\pi\over 2}-\eta\ln (2kr)+\delta_k)\\
\sqrt{{\epsilon-m_ec^2\over 2\epsilon}}
\cos(kr-l{\pi\over 2}-\eta\ln (2kr)+\delta_k)\end{array}\right)
\end{eqnarray}
Here 
$c$ is the speed of the light,  $m_e/\epsilon$ are the electron mass/energy, 
$k={p/\hbar}$ is the electron wave number, $\eta=Ze^2/\hbar$ (with $Z=\pm Z$ for $\beta^{\mp}$ decays),  $v$ is the Sommerfeld parameter, $\delta_\kappa$ is the phase shift and $V$ is the Coulomb interaction energy between the electron and the daughter nucleus. 
For the continuum spectrum, the radial function is normalized to the
asymptotic form of the Coulomb function. The phase shifts are obtained by
matching the inner numerical solution to the analytic function.

The bound states w.f. for the electron
\begin{equation}
\Psi^b_{\epsilon_n\kappa\mu}(r) = \left(\begin{array}{l}g_{n,\kappa}(r)\chi^\mu_\kappa\\if_{n,\kappa}(r)\chi^\mu_{-\kappa}\end{array}\right)
\end{equation}
are solutions of the Dirac equation (\ref{dirac}) and correspond
to the eigenvalues $\epsilon_{n}$ ($n$ is the radial
quantum number). The quantum number $\kappa$ is related to
the total angular momentum $j_\kappa=\mid\kappa\mid-1/2$.   These wave functions are normalized such that
\begin{equation}
\int_0^\infty [g^2_{n,\kappa}(r)+f^2_{n,\kappa}(r)]dr=1.
\end{equation}
An asymptotic solution is obtained by means of the WKB approximation 
and by considering that the potential $V$ is
negligible small:
\begin{equation}
{f_{n,\kappa}\over g_{n,\kappa}}={c\hbar\over \epsilon+m_ec^2}\left({g'_{n,\kappa}\over g_{n,\kappa}}+{\kappa\over r}\right)
\end{equation}
where
\begin{equation}
{g'_{n,\kappa}\over g_{n,\kappa}}=-{1\over 2}\mu'\mu^{-1}-\mu
\end{equation}
with
\begin{equation}
\mu=\left[{\epsilon+m_ec^2\over \hbar^2 c^2}(V-\epsilon+m_ec^2)+{\kappa^2\over r^2}\right]^{1/2}.
\end{equation}
In our calculations we  use $n$=0 and $n$=1 number of nodes, for
the orbitals $1s_{1/2}$ and $2s_{1/2}$, $\kappa$ being -1.
The eigenvalues of the discrete spectrum are obtained by matching
two numerical solutions of the Dirac equation: the inverse solution
that starts from the asymptotic conditions and the direct one that starts at $r$=0.

\subsection{The Coulomb potential}

The nuclear size corrections are usually taken into account by considering an unscreened potential $V$ obtained for a uniform charge 
distribution in a sphere of radius $R_A$ \cite{DOI83}, \cite{KI12}:
\begin{equation}
V(Z,r)=\left\{\begin{array}{ll}-{Z{\alpha\hbar c}\over r}, & r \ge R_A,\\
-Z(\alpha\hbar c)\left({3-(r/R_A)^2\over 2R_A}\right), & r<R_A, \end{array}
\right.
\label{cd}
\end{equation}
where generalized atomic units are used. The values of
$\hbar$, of the electron
charge $e$ and its mass $m_e$ are considered as unity. The energy
unit is $E_0=me^4/h^4=27.2114$ eV, the Bohr radius is 
$a_0=\hbar^2/m_ee^2$=0.529177 $\AA$  and the speed of the light
in vacuum is $c$=137.036 (the inverse of the fine structure constant).

In this work we take into account the influence of the nuclear structure by using a potential $V(r)$ derived from a realistic proton density distribution
in the nucleus. This is done by solving the Schr\"odinger equation for a Woods-Saxon potential. In this case:
\begin{equation}
V(Z,r)=\alpha\hbar c\int{\rho_e(\vec{r'})\over \mid \vec{r}-\vec{r'}\mid}
d\vec{r'}
\label{vpot}
\end{equation}
where the charge density is
\begin{equation}
\rho_e({\vec{r}})=\sum_i (2j_i+1)v_i^2 \mid\Psi_i(\vec{r})\mid^2
\end{equation}
$\Psi_i$ is the proton (Woods-Saxon) w. f.  of the spherical single particle state
$i$ and $v_i$ is its occupation amplitude. The factor $(2j_i+1)$ reflects the
spin degeneracy.

\begin{figure}
\includegraphics[width=0.5\textwidth]{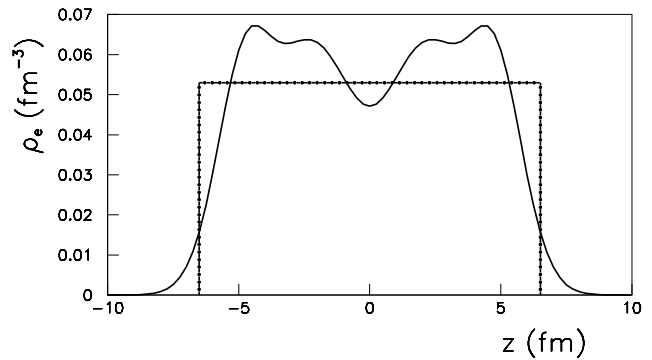}
\caption{ 
 Profile of the realistic proton density $\rho_e$ for $^{150}$Sm 
(thick line)
compared with that given with the constant density approximation (dot-dashed line).
\label{dens}}
\end{figure}

The difference between the behavior of the constant charge density $\rho_e$
and the realistic charge density is displayed in Fig. \ref{dens} for
the daughter nucleus $^{150}$Sm. 
We computed the Coulomb potential with formula (\ref{vpot}).
In this case, the differences given
by the charge densities are translated in a shift of 0.5 MeV energy
in the potential at $r=0$. This difference in energy vanishes when
$r$ increases, but is able to affect the values of the w.f.

The screening effect is taken into account by multiplying the expression of $V(r)$ with a function $\phi(r)$, which is the solution of the Thomas Fermi equation:
$d^2\phi/dx^2 = \phi^{3/2}/\sqrt x$, with $x=r/b$, $b\approx 0.8853a_0Z^{-1/3}$ and $a_0$ = Bohr radius. It is
calculated within the Majorana method \cite{sal}. The screening effect is taken into account in the same manner as in Ref. \cite{KI12}. 
The modality in which the screening function modifies
the Coulomb potential depends on the specific mechanism and its
boundary conditions.

\begin{figure}
\includegraphics[width=0.5\textwidth]{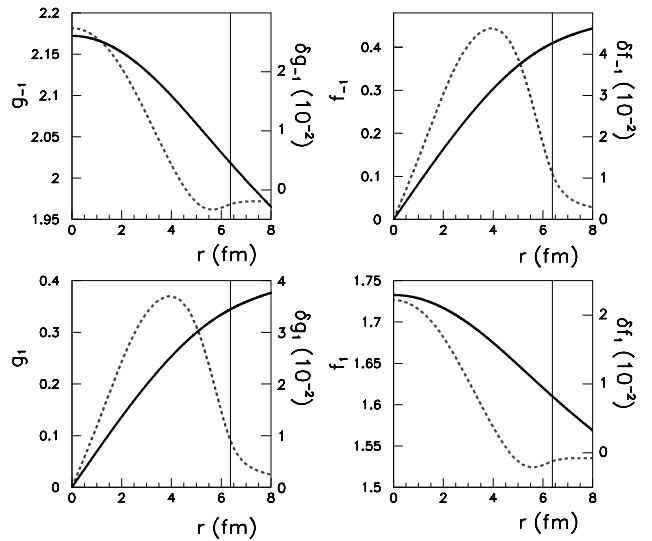}
\caption{The absolute values of the electron phase space parameter
$f_i(\epsilon,r)$ and $g_i(\epsilon,r)$
($i$=-1,1) are plotted with a full line for the daughter nucleus
$^{150}$Sm at an energy $\epsilon$=2 MeV
as function of the distance $r$. The deviations that arise when only the constant charge
distribution is taken into account, $\delta g_i(\epsilon,r)$ 
and $\delta f_i(\epsilon,r)$, are displayed as well
with a dashed line and their scales are on the right. 
A thin line locates the nuclear surface $R=r_0\times A^{1/3}$.}
\label{furi}
\end{figure}

In the case of the $\beta^-\beta^-$ process, the potential used to obtain the electron w.f. is:
\begin{equation}
rV_{\beta^-\beta^-}(Z,r)=(rV(Z,r)+2)\times\phi(r)-2
\end{equation}
to take into account the fact that DBD releases a final 
positive ion with charge +2. Here, the product $\alpha\hbar c$=1, for
atomic units. In this case, the charge
number $Z=Z_0+2$ corresponds to the daughter nucleus,
$Z_0$ being the charge number of the parent nucleus.
In the case of the $\beta^+\beta^+$ process,
the potential used to obtain the electron wave functions
is:
\begin{equation}
rV_{\beta^+\beta^+}(Z,r)=(rV(Z,r)-2)\times\phi(r)+2
\end{equation} 
where the final configuration is characterized by an ion with charge -2.
In this case, the daughter nucleus has the charge number $Z=Z_0-2$.
In both approaches, at $r$=0 the potential is unscreened because
$\phi(0)$=1. Asymptotically $\phi(r)$ tends to 0 and we are left with the
charge number of the final system. 

In the case of the $EC$ process, the potential used to obtain the electron w.f. reads:
\begin{equation}
rV_{EC}(Z,r)=(rV(Z,r)+1)\times\phi(r)-1
\end{equation}
\noindent 
and the charge number $Z=Z_0$ corresponds to the parent nucleus. In the case of the 
$\beta^+$ process, the potential used to obtain the positron wave functions reads:
\begin{equation}
rV_{\beta^+}(Z,r)=(rV(Z,r)-1)\times\phi(r)+1
\end{equation}
to take into account that in the final 
configuration we have an ion with charge -1.
 In this case the daughter nucleus has the charge number $Z=Z_0-1$.
In the case of the ECEC process, the potential used to obtain the electron wave functions is:
\begin{equation}
V_{ECEC}(Z,r)=V(Z,r)\times\phi(r)
\end{equation}
and $Z=Z_0$, the final system being neutral.

The  solutions of the Dirac equation for free states at $\epsilon$=2 MeV in the case
of $^{150}$Sm    are plotted in Fig. \ref{furi}. With a dashed line
we plotted the deviations $\delta g_i(\epsilon,r)=g_i(\epsilon,r)-\tilde{g}_i(\epsilon,r)$ and
$\delta f_i(\epsilon,r)=f_i(\epsilon,r)-\tilde{f}_i(\epsilon,r)$ ($i$=-1,1) that are due to the
constant density approximation. The quantities $g_i$ and $f_i$ are
obtained for a realistic charge distribution while $\tilde{g}_i$ and
$\tilde{f}_i$ are calculated by considering a constant charge density
in the nucleus. The deviations exhibit  an oscillatory behavior leading 
to uncertainties on the nuclear surface. The nuclear surface is marked with a vertical line
in the figure.

\subsection{Calculation of the phase space factors}

\subsubsection{Double Electron and Double Positron decay modes}

To compute the PSF, we have to obtain the electron phase factors $f_{jk}^{(0)}$

\begin{equation}
f_{11}^{(0)}=\mid f^{-1-1}\mid^2+\mid f_{11}\mid^2+\mid f^{-1}_{~~~1}\mid^2
+\mid f^{~-1}_1\mid^2
\end{equation}
with
\begin{eqnarray}
f^{-1-1}=g_{-1}(\epsilon_1)g_{-1}(\epsilon_2)~;~f_{11}=f_1(\epsilon_1)f_1(\epsilon_2),\\
f^{-1}_{~~~1}=g_{-1}(\epsilon_1)f_1(\epsilon_2)~;~ f_{1}^{~-1}=f_1(\epsilon_1)g_1(\epsilon_2)
\end{eqnarray}
\noindent
from  the solutions of the Dirac equation by considering s-wave states and neglecting the neutrino mass.
The values of the $f$ and $g$ functions are approximated with the solutions on the nuclear surface (the method I from \cite{KI12}).
\begin{eqnarray}
g_{-1}(\epsilon)=g_{-1}(\epsilon,R_A)~;~ f_1(\epsilon)=f_1(\epsilon,R_A)
\end{eqnarray}
\noindent
where $R_A=1.2A^{1/3}$ fm.
For the $2\nu\beta\beta$ decay mode and for transitions to ground states, the PSF expression reads:
\begin{eqnarray}
G^{\beta\beta}_{2\nu}(0^+\rightarrow0^+)={2\tilde{A}^2\over 3\ln 2 g_A^4(m_ec^2)^2}
\int_{m_ec^2}^{Q^{\beta\beta}+m_ec^2} d\epsilon_1 \nonumber\\
\times \int_{m_ec^2}^{Q^{\beta\beta}+2m_ec^2-\epsilon_1} d\epsilon_2
\int_{0}^{Q^{\beta\beta}+2mc_e^2-\epsilon_1-\epsilon_2} 
   d\omega_1 \nonumber\\ 
\times f_{11}^{(0)}w_{2\nu}(\langle K_N\rangle^2+\langle L_N\rangle^2+
\langle K_N\rangle \langle L_N\rangle) 
\label{fomg}
\end{eqnarray}

\noindent
where 
$Q^{\beta\beta}=M(A,Z_0)-M(A,Z_0-2)-4m_ec^2$ is  the kinetic energy released in the process.
$\langle K_N\rangle$, $\langle L_N\rangle$ are expressions that depend on the electron/positron 
$(\epsilon_{1,2})$ and neutrino ($\omega_{1,2}$) energies, and on the g.s. energy of 
the parent nucleus and on the excited states energy of the intermediate nucleus \cite{HS84}-\cite{KI12}.

\begin{equation}
\langle K_N\rangle=
{1\over \epsilon_1+\omega_1+\langle E_N\rangle-E_I}+
{1\over \epsilon_2+\omega_2+\langle E_N\rangle-E_I}
\label{ie1}
\end{equation}
\begin{equation}
\label{ie2}
\langle L_N\rangle=
{1\over \epsilon_1+\omega_2+\langle E_N\rangle-E_I}+
{1\over \epsilon_2+\omega_1+\langle E_N\rangle-E_I}
\end{equation}
Here, the difference in energy in the denominator can be
obtained from the approximation
$\tilde{A}^2=[W_0/2+\langle E_N\rangle -E_I]^2$, where
$\tilde{A}=1.12A^{1/2}$ (in MeV) gives the energy of the giant Gamow-Teller resonance in the intermediate nucleus. The quantity $W_0$ is
related to the $Q$ value of the process and 
\begin{equation}
w_{2\nu}={g_A^4(G\cos\theta_C)^4\over 64\pi^7\hbar}
\omega_1^2\omega_2^2(p_1c)(p_2c)\epsilon_1\epsilon_2.
\end{equation}
\noindent
where $\omega_1$ and $\omega_2=Q^{\beta\beta}-\epsilon_1-\epsilon_2-\omega_1+2m_ec^2$ are
the neutrino energies.
The PSF are finally renormalized to the electron rest energy and are reported in $[yr^{-1}]$.

The PSF for the $2\nu\beta\beta$ decay mode and for transitions to excited $0^+_1$ states is calculated with a formula similar to (\ref{fomg}),
 but replacing $Q^{\beta\beta}$ by $Q(0^+_1) = Q^{\beta\beta} - E_x(0^+_1)$, which is the kinetic energy released in this transition. $E_x(0^+_1)$ is the energy of the excited $0^+_1$ state of the daughter nucleus $x$.     

For the $2\nu\beta\beta$ decay mode and for transitions to excited $2^+_1$ states, the PSF formula reads \cite{SC98}-\cite{SM13}:

\begin{eqnarray}
G^{\beta\beta}_{2\nu}(0^+\rightarrow2^+_1)={2\tilde{A}^6\over \ln 2 g_A^4(m_ec^2)^2}
\int_{m_ec^2}^{Q^{\beta\beta}{(2^+_1)}+m_ec^2} d\epsilon_1 
\nonumber\\
\times \int_{m_ec^2}^{Q^{\beta\beta}{(2^+_1)}+2m_ec^2-\epsilon_1} d\epsilon_2\nonumber \int_{0}^{Q^{\beta\beta}{(2^+_1)}+2mc_e^2-\epsilon_1-\epsilon_2} d\omega_1 \nonumber \\
\times
f_{11}^{(0)}w_{2\nu}(\langle K_N\rangle-\langle L_N\rangle)^2 
\end{eqnarray}

where $Q(2^+_1) = Q - E_x(2^+_1)$.

For the $0\nu\beta\beta$ decay and for transitions to g.s. the PSF reads:
\begin{equation}
\label{int1}
G^{\beta\beta}_{0\nu}(0^+\rightarrow0^+) = {2\over{4g^4_AR_A^2 \ln2}}\int_{m_ec^2}^{Q^{\beta\beta}+m_ec^2}
f_{11}^{(0)}w_{0\nu}d\epsilon_1
\end{equation}

where $R_A=1.2A^{1/3}$ is the nuclear radius and
\begin{equation}
w_{0\nu }={g_A^4(G\cos\theta_C)^4\over 16\pi^5}
(m_ec^2)^2 (\hbar c^2) (p_1c) (p_2c) \epsilon_1\epsilon_2
\end{equation}
In our calculations, the Fermi constant is $G=1.16637\times10^{-5}$ GeV$^{-2}$ 
and $\cos\theta_C$=0.9737.
In Eq.(\ref{fomg}) it is convenient to redefine the PSF by a renormalization that eliminates the constant $g_A$ and correlate (by dividing by $4R_A^2$) the dimension of $G_{0\nu}$ with the NME which are dimensionless. Thus, the PSF are also reported in $[yr^{-1}]$. A similar expression is employed in the PSF calculation for the transitions to excited $0^+_1$ states, but replacing $Q^{\beta\beta}$ by  $Q^{\beta\beta}(0^+_1)$.         
The formula used for the calculation of the PSF for $2\nu\beta^+\beta^+$ is similar to that used for $2\nu\beta^-\beta^-$ decay, but $\epsilon_{1,2}$ are now the positron energies. Also, we use the same approximations as described above,
to evaluate the radial positron w. f. ($g$ and $f$) at the nuclear surface and replace the excitation energy $E_N$ in the intermediate odd-odd nucleus by a suitable average energy.

\subsubsection{The EC$\beta^+$ case}

For the $EC\beta^+$ decays the energy released in the process is $Q^{EC\beta}=M(A,Z_0)-M(A,Z_0-2)-2mc_e^2$.
If the numerical solutions  of the Dirac equation are obtained 
in Bohr units $a_0$, the probability that an electron is found on the surface
of a nucleus of radius $R_A$ can be defined as:
\begin{equation}
B^2_{n,\kappa}={1\over 4\pi (m_ec^2)^3}\left({\hbar c\over a_0}\right)^2
\left({a_0\over R_A}\right)^2
[g_{n,\kappa}^2(R_A)+f_{n,\kappa}^2(R_A)]
\end{equation}
The PSF expression for $2\nu\beta\beta$ decay modes is
\begin{eqnarray}
G^{EC\beta^+}_{2\nu}=
{2A^2\over 3\ln2}{(G\cos\theta)^4\over 16\pi^5\hbar}
(m_ec^2)\sum_{i=0,1}B^2_{i,-1} \nonumber \\
\times \int_{m_ec^2}^{Q^{EC\beta}+\epsilon_{i,-1}+m_ec^2}
\int_0^{Q^{EC\beta}+\epsilon_{i,-1}-\epsilon_p+m_ec^2}
\nonumber \\
\times [g_{-1}^2(\epsilon_p)+f_{1}^2(\epsilon_p)]\nonumber \\
\times
(\langle K_N\rangle^2+\langle L_N\rangle^2+
\langle K_N\rangle\langle L_N\rangle)\omega_1^2\omega_2^2
p_pc\epsilon_pd\omega_1d\epsilon_p
\end{eqnarray}
where $\epsilon_{n,\kappa}$ are the binding energies
of the electron while $p_p$ and $\epsilon_p$ are
the linear momentum and the energy of the positron.
Here, the expressions for 
$\langle K_N\rangle$ and $\langle L_N\rangle$ are similar to those from Eqs. (\ref{ie1})-(\ref{ie2}), but where 
$\epsilon_1$ is replaced by $\epsilon^c_{i,-1}=m_ec^2-\epsilon_{i,-1}$, the energy of the captured electron 
and $\epsilon_2$ is replaced by $\epsilon_p$, the energy of the emitted positron.
For the $0\nu\beta\beta$ decay process the PSF expression is:
\begin{eqnarray}
G^{EC\beta^+}_{0\nu}={1\over 4 R_A^2}{2\over \ln2}
{(G\cos\theta)^4\over 4\pi^3}(\hbar c^2)(m_ec^2)^5
\nonumber \\ \times
\sum_{i=0,1}B^2_{i,-1}
[g_{-1}^2(\epsilon_{p,i})+f_{1}^2(\epsilon_{p,i})]p_pc\epsilon_p
\end{eqnarray}
where $\epsilon_{p,i}$ denotes the maximal value of the positron associated to the state $i$.

\subsubsection{The $2\nu ECEC$ case}

The PSF expression is defined as:
\begin{eqnarray}
G_{2\nu}^{ECEC}={2\tilde A^2\over 3\ln 2}
{(G\cos\theta)^4\over 16\pi^3\hbar}
(mc_e^2)^4
\nonumber \\ \times
\sum_{i,j=0,1}B^2_{i,-1}B^2_{j,-1}
\int_0^{Q^{ECEC}+\epsilon_{i,-1}+\epsilon_{j,-1}}
\nonumber \\ \times
(\langle K_N\rangle^2+\langle L_N\rangle^2+
\langle K_N\rangle\langle L_N\rangle)\omega_1^2\omega_2^2d\omega_1
\end{eqnarray}
\noindent
where $Q^{ECEC}=M(A,Z_0)-M(A,Z_0-2)$ is the energy released in the process.
The expressions for 
$\langle K_N\rangle$ and $\langle L_N\rangle$ are similar to those from Eqs. (\ref{ie1})-(\ref{ie2}), but where 
$\epsilon_{(1,2)}$ are replaced by $\epsilon^c_{(i,j)-1}=m_ec^2-\epsilon_{(i,j)-1}$, the energies of the captured electrons.

\section{Numerical details}
\label{3}

The numerical solutions of the Dirac equation were obtained by using the
power series method of Ref. \cite{buh}. We built up a code that use
a numerical algorithm
similar to that described in Ref. \cite{sal1}, the normalization to unity for free states  being
done as indicated in Ref. \cite{sal2}.

In the numerical procedure, the potential energy 
as function of the distance $r$ is approximated
with a spline cubic function that interpolates
values defined by an increment $x$.
The radial w. f. is expanded as an infinite power series
that depends on the increment and the coefficients
of the spline function. Therefore, the values
of the w. f. are calculated step by step in the
mesh points. The accuracy of the solutions depends
on the increment and the number of terms in the series
expansion. We used an increment interval of 10$^{-4}$ fm
and at least 100 terms in the series expansion. These values
exceed the convergence criteria of Ref. \cite{sal1}.
At very large distances, the behavior of the w. f.
 must resemble to that of the Coulomb function. This last
condition provides a way to renormalize the amplitude to unity
 and to determine the phase shift.
For discrete states, the asymptotic behavior of the w.f. gives a boundary for the inverse solutions. The eigenvalue
is obtained when the direct
solutions and the inverse ones match together.

In order to find the bound states of the electron, a procedure
which differ from that given in Ref. \cite{sal1} is used. We start to
compute numerical solution of the Dirac equation for a total
energy close to $m_ec^2$, and we span an interval of 0.3 MeV under
this energy in  steps of 0.0002 MeV. In this range of energies, all the possible
bound solutions are found. In the Ref. \cite{sal1}, the interval
of the allowed solutions is fixed by initial conditions, 
but its lower limit is
an approximate one and sometimes the equations cannot be solved
numerically.

For the PSF computation, 
all integrals are performed accurately with Gauss-Legendre quadrature in 32 points.
We calculated up to 49 values of the radial functions in the Q value
energy interval, values that are interpolated with spline functions.
In our calculations we used up-dated values of $Q$ reported recently in  Ref. \cite{audi12}. 

\section{Results and discussion}
\label{4}

Our results are tabulated in tables \ref{tab1} - \ref{tab6}.
For comparison with other previous results we refer mainly to those reported by KI \cite{KI12}-\cite{KI13} because both methods (KI and ours) calculate the PSF more precisely than other previous ones, i.e. using exact electron w.f. obtained by solving numerically the Dirac equation with inclusion of the finite nuclear size and electron screening effects. In addition, in our calculations we use a Coulomb potential derived from a realistic proton density distribution in the daughter/parent nucleus, improve the numerical precision of the routines that we developed both to solve the Dirac equation and to integrate the PSF values, and use updated $Q$-values \cite{audi12}. These additional ingredients lead in many cases to differences as compared with KI's results, that we qualitatively discuss in the following. For the $\beta^-\beta^-$ decay mode our results are in good agreement with KI's results \cite{KI12}, the differences between their values and ours are within 3\% for $G_{0\nu}$ and for the transitions to final g.s., and within 10\% for transitions to excited $0^+_1$ states, with two exceptions, $^{110}Te$ and $^{130}Te$ where the differences are within 18\% and 38\%, respectively. For $G_{2\nu}$ the differences are within 7\%, except the nuclei $^{128}Te$ (20\%) and $^{238}U$ (our result is 7 times larger than the KI's one). We mention that $\beta^-\beta^-$ decay modes are experimentally the most interesting DBD because the $2\nu$ decay mode is already measured for eleven nuclei for transitions to g.s., and for two nuclei for transitions to excited $0^+_1$. Also, the $0\nu$ channel is intensively searched for checking the LNC and extracting information about neutrino properties. Although not yet measured, the $\beta^-\beta^-$ decays to excited $2^{+}_1$ states are interesting as well, to probe alternative mechanisms for $0\nu$ decay  to the most common one, i.e. by exchange of a virtual light neutrino between two nucleons inside the nucleus. For these transitions KI have not reported PSF values. Comparing our PSF values with older results \cite{DOI83}, \cite{SC98}, we found differences in the range (17-86)\%, for both $G_{0\nu}$ and $G_{2\nu}$, unless $^{76}Ge$, $^{116}Cd$ and  $^{128}Te$ where our results are several times larger. For the $\beta^+\beta^+$ decay modes our PSF values are in agreement with KI's results within (3-12)\%  for $G_{0\nu}$ and within (6-23)\% for $G_{2\nu}$. For $EC\beta^+$ decay mode the differences between KI's results and ours are within 35\% both for $G_{0\nu}$ and $G_{2\nu}$ for the most part of the isotopes, unless a few cases where PSF values amount to about 70\%. There are, however, some heavier nuclei where our results differ by  factors of 5-10 from other previous results. Finally, for the $ECEC$ decay mode the differences between our $G_{2\nu}$ values and the KI ones, are within (30-55)\%, unless the heavier nuclei where the differences are significantly larger. It is worth to mention that we got significant differences as compared with KI's results especially for the decays where the $Q$-values are small. In these cases the numerical precision in the PSF integration is important and must be treated carefully. The use of a realistic Coulomb potential can also lead to differences between KI and our results, especially for double-positron emitting and/or positron emitting  electron(s) captured decays. The use of recently reported $Q$ values \cite{audi12} may bring other differences between our results and previous ones. For example,  we first computed the PSF with the same $Q$-values used by KI and then with the  $Q$-values taken from \cite{audi12} and got differences of (3-4)\% between the two sets of results. We mention that, in our opinion, differences of 10\% or more in the PSF values are significant for precise DBD calculations/estimations, hence the evaluation of the PSF for many DBD decay modes and/or transitions is still a challenge and it is necessary to investigate it thoroughly.

\section{Conclusions}
\label{5}
We recalculated with a more precise method the PSF involved in the $\beta^-\beta^-$, $\beta^+\beta^+$, 
$EC\beta^+$ and $ECEC$ DBD modes. In addition to the previous (most accurate) method \cite{KI12}-\cite{KI13}, we use a Coulomb potential derived from a realistic proton density distribution in nucleus, improve the precision of our numerical routines used to solve the Dirac equations and to integrate the PSF expressions, and  use $Q$ values taken from a recent mass evaluation \cite{audi12}. These ingredients lead in many cases to significant corrections to the present available PSF values, which are discussed as well. Accurate values of the PSF (besides the NME) are necessary ingredients in the DBD study, both for theorists to improve the lifetimes predictions and constraint the neutrino parameters, and for experimentalists, to plan their set-ups. 
 \\

{\it \bf Acknowledgments}. This work was supported by a grant of the Romanian Ministry of National Education, CNCS UEFISCDI, project PCE-2011-3-0318, Contract no. 58/28.10/2011.




\newpage
\begin{table*}
\caption{PSF for $\beta^-\beta^-$ decays to final g.s.}
\label{tab1}
\begin{tabular}{|c|c|c|c|}
\hline
\textit{Nucleus} & \textit{Q}$^{\beta^-\beta^-}_{g.s.}$ (MeV) & \textit{G}$_{2\nu}^{\beta^-\beta^-}(g.s.)$ (10$^{-21}$ yr$^{-1}$)&\textit{G}$_{0\nu}^{\beta^-\beta^-}(g.s.)$ (10$^{-15}$ yr$^{-1}$) \\ & & &  \\
\hline
$^{48}$Ca	&4.267	&15536	&24.65		\\
$^{76}$Ge	&2.039	&46.47	&2.372		\\
$^{82}$Se	&2.996	&1573	&10.14		\\
$^{96}$Zr	&3.349	&6744 	&20.48 		\\
$^{100}$Mo	&3.034	&3231	&15.84		\\
$^{110}$Pd	&2.017	&132.5	&4.915		\\
$^{116}$Cd	&2.813	&2688	&16.62		\\
$^{128}$Te	&0.8665	&0.2149	&0.5783		\\
$^{130}$Te	&2.528	&1442	&14.24		\\
$^{136}$Xe	&2.458	&1332	&14.54		\\
$^{150}$Nd	&3.371	&35397	&61.94		\\
$^{238}$U	&1.144	&98.51	&32.53		\\  \hline
\end{tabular}
\end{table*}

\newpage
\newpage

\begin{table*}
\label{tab2}
\caption{PSF for $\beta^-\beta^-$ decays to final excited 0$^{+}_{1}$ states }
\begin{tabular}{|c|c|c|c|}
\hline
\textit{Nucleus} & \textit{Q}$^{\beta^-\beta^-}_{0^{+}_{1}}$ (MeV) & \textit{G}$_{2\nu}^{\beta^-\beta^-}(0^{+}_{1})$ (10$^{-21}$ yr$^{-1}$)&\textit{G}$_{0\nu}^{\beta^-\beta^-}(0^{+}_{1})$ (10$^{-15}$ yr$^{-1}$) \\ & & & \\
\hline
$^{48}$Ca	&	1.270	&0.3518		&0.3041		\\
$^{76}$Ge	&	0.9171	&0.06129	&0.1932		\\	
$^{82}$Se	&	1.508	&4.170		&0.9440		\\
$^{96}$Zr	&	2.201	&169.4		&4.594		\\
$^{100}$Mo	&	1.904	&57.08		&3.168		\\
$^{110}$Pd	&	0.5472	&3.280$\times 10^{-3}$	&0.1223		\\	
$^{116}$Cd	&	1.056	&0.7590		&0.7585		\\
$^{130}$Te	&	0.7335	&0.05460	&0.3651		\\	
$^{136}$Xe	&	0.8790	&0.2823		&0.6746		\\
$^{150}$Nd	&	2.631	&4116		&26.96		\\
$^{238}$U	&	0.2032	&1.491$\times10^{-4}$	&0.8229		\\ \hline	
\end{tabular}
\end{table*}



\newpage
\begin{table*}
\label{tab3}
\caption{PSF for $\beta^-\beta^-$ decays to final excited 2$^{+}_{1}$ states}
\begin{tabular}{|c|c|c|c|}
\hline
\textit{Nucleus} & \textit{Q}$^{\beta^-\beta^-}_{2^{+}_{1}}$ (MeV) & \textit{G}$_{2\nu}^{\beta^-\beta^-}(2^{+}_{1})$ (10$^{-21}$ yr$^{-1}$)&\textit{G}$_{2\nu}^{\beta^-\beta^-}(2^{+}_{1})$ (10$^{-15}$ yr$^{-1}$) \\ & & & \\
\hline
$^{48}$Ca		&	3.284	&1374		&8.741		\\
$^{76}$Ge		&	1.480	&3.083		&0.8154		\\
$^{82}$Se		&	2.219	&111.4		&3.445		\\
$^{96}$Zr		&	2.571	&654.5		&7.885		\\
$^{100}$Mo		&	2.494	&572.1		&7.882		\\
$^{110}$Pd		&	1.359	&5.017		&1.415		\\
$^{116}$Cd		&	1.520	&14.10		&2.182		\\
$^{128}$Te		&	0.4255	&8.215$\times 10^{-4}$	&0.1063		\\	
$^{130}$Te		&	1.990	&192.9		&6.486		\\
$^{136}$Xe		&	1.640	&42.80		&3.932		\\
$^{150}$Nd		&	3.037	&14084		&43.30		\\
$^{238}$U		&	1.099	&73.77		&29.64		\\  \hline
\end{tabular}
\end{table*}

\newpage


\begin{table*}
\label{tab4}
\caption{PSF for $\beta^+\beta^+$ decay mode}
\begin{tabular}{|c|c|c|c|}
\hline
\textit{Nucleus} & \textit{Q}$^{\beta^+\beta^+}$  (MeV) & \textit{G}$_{2\nu}^{\beta^+\beta^+}$ (10$^{-29}$ yr$^{-1}$)&\textit{G}$_{0\nu}^{\beta^+\beta^+}$ (10$^{-20}$ yr$^{-1}$) \\ & & & \\
\hline
$^{78}$Kr	&0.8023	&	9159	&	243.2	\\
$^{96}$Ru	&0.6706	&	942.3	&	80.98	\\
$^{106}$Cd	&0.7314	&	1794	&	91.75	\\
$^{124}$Xe	&0.8203	&	4261	&	107.8	\\
$^{130}$Ba	&0.5748	&	91.54	&	23.82	\\
$^{136}$Ce	&0.3345	&	0.2053	&	2.126	\\  & & & \\ \hline
\end{tabular}
\end{table*}





\newpage

\begin{table*}
\caption{PSF for $EC\beta^+$ decay mode }
\label{tab5}
\begin{tabular}{|c|c|c|c|c|c|}
\hline
\textit{Nucleus} & \textit{Q}$^{EC\beta^+}$  (MeV) & $\epsilon_{0,-1}$   (keV)& $\epsilon_{1,-1}$   (keV)& \textit{G}$_{2\nu}^{EC\beta^+}$ (10$^{-24}$ yr$^{-1}$)&\textit{G}$_{0\nu}^{EC\beta^+}$ (10$^{-18}$ yr$^{-1}$) \\ & & & & & \\
\hline
$^{78}$Kr	&	1.824	&	17.7	&	3.1	&	338.1	&	6.901	\\
$^{96}$Ru	&	1.693	&	26.2	&	4.9	&	372.1	&	11.30	\\
$^{106}$Cd	&	1.753	&	31.1	&	5.9	&	741.2	&	15.39	\\
$^{124}$Xe	&	1.842	&	39.4	&	7.8	&	1235	&	17.10	\\
$^{130}$Ba	&	1.597	&	42.4	&	8.5	&	740.1	&	22.89	\\
$^{136}$Ce	&	1.357	&	45.6	&	9.2	&	255.6	&	21.45	\\  & & & & & \\
$^{50}$Cr	&	0.1469	&	8.3	&	1.2	&	1.62$\times10^{-6}$	&	0.376 	\\
$^{58}$Ni	&	0.9043	&	11.1	&	1.2	&	1.037 	&	1.623	\\
$^{64}$Zn	&	0.07269	&	12.6	&	2.0	&	2.207$\times10^{-8}$	&	0.546 	\\
$^{ 74 }$Se	&	0.1872	&	16.0	&	2.7	&	2.197$\times10^{-5}$	&	0.9558	\\
$^{ 84 }$Sr	&	0.7677	&	19.7	&	3.5	&	0.7821	&	3.067	\\
$^{ 92 }$Mo	&	0.6298	&	23.9	&	4.4	&	0.245 	&	3.526	\\
$^{ 102 }$Pd	&	0.1499	&	28.6	&	5.4	&	1.401$\times10^{-5}$	&	2.101	\\
$^{ 112 }$Sn	&	0.8978	&	33.7	&	6.5	&	6.313	&	8.405	\\
$^{ 120 }$Te	&	0.7082	&	36.5	&	7.1	&	1.406	&	7.609	\\
$^{ 144 }$Sm	&	0.7604	&	52.3	&	10.7	&	3.56 	&	10.38	\\
$^{ 156 }$Dy	&	0.9840	&	59.6	&	12.4	&	48.1 	&	26.01	\\
$^{ 162 }$Er	&	0.8250	&	63.4	&	13.3	&	15.41	&	25.03	\\
$^{ 168 }$Yb	&	0.3873	&	67.4	&	14.0	&	9.051$\times10^{-3}$	&	16.48	\\
$^{ 174 }$Hf	&	0.07689	&	71.6	&	15.2	&	7.512$\times10^{-6}$	&	10.81 	\\
$^{ 184 }$Os	&	0.4289	&	80.4	&	17.3	&	0.321               	&	26.21	\\
$^{ 190 }$Pt	&	0.3625	&	85.2	&	18.2	&	0.127               	&	27.61	\\ & & & & & \\ \hline
\end{tabular}
\end{table*}




\newpage

\begin{table}
\caption{PSF for $2\nu ECEC$ decay mode}
\label{tab6}
\begin{tabular}{|c|c|c|c|c|}
\hline\\
\textit{Nucleus} & \textit{Q}$^{ECEC}$  (MeV) & $\epsilon_{0,-1}$  (keV)& $\epsilon_{1,-1}$  (keV)& \textit{G}$_{2\nu}^{ECEC}$ (10$^{-24}$ yr$^{-1}$) \\ & & & & \\
\hline\\

$^{78}$Kr	&	2.846	&	17.7	&	3.1	&	409.8		\\
$^{96}$Ru	&	2.715	&	26.2	&	4.9	&	1450		\\
$^{106}$Cd	&	2.775	&	31.1	&	5.9	&	4299		\\
$^{124}$Xe	&	2.864	&	39.4	&	7.8	&	15096	\\
$^{130}$Ba	&	2.619	&	42.4	&	8.5	&	14773	\\
$^{136}$Ce	&	2.379	&	45.6	&	9.2	&	12223	\\ 
& & & &  \\
$^{50}$Cr	&	1.169	&	8.3	&	1.2	&	0.2376		\\
$^{58}$Ni	&	1.926	&	11.1	&	1.2	&	9.903		\\
$^{64}$Zn	&	1.095	&	12.6	&	2.0	&	1.0305		\\
$^{ 74 }$Se	&	1.209	&	16.0	&	2.7	&	3.409		\\
$^{ 84 }$Sr	&	1.790	&	19.7	&	3.5	&	64.62		\\
$^{ 92 }$Mo	&	1.652	&	23.9	&	4.4	&	82.32		\\
$^{ 102 }$Pd	&	1.172	&	28.6	&	5.4	&	42.09		\\
$^{ 112 }$Sn	&	1.920	&	33.7	&	6.5	&	869.7		\\
$^{ 120 }$Te	&	1.730	&	36.5	&	7.1	&	840.3		\\
$^{ 144 }$Sm	&	1.782	&	52.3	&	10.7	&	6436		\\
$^{ 156 }$Dy	&	2.006	&	59.6	&	12.4	&	22078	\\
$^{ 162 }$Er	&	1.847	&	63.4	&	13.3	&	20085	\\
$^{ 168 }$Yb	&	1.409	&	67.4	&	14.0	&	7872		\\
$^{ 174 }$Hf	&	1.099	&	71.6	&	15.2	&	3432		\\
$^{ 184 }$Os	&	1.451	&	80.4	&	17.3	&	24222	\\
$^{ 190 }$Pt	&	1.384	&	85.2	&	18.2	&	28153	\\ 
&  & & &  \\
$^{ 36 }$Ar	&	0.4326	&	5.0	&	1.2	&	2.902$\times 10^{-4}$	\\	
$^{ 40 }$Ca	&	0.1935	&	5.9	&	1.2	&	1.021$\times 10^{-5}$	\\	
$^{ 54 }$Fe	&	0.6798	&	9.7	&	1.4	&	0.03021	\\	
$^{ 108 }$Cd	&	0.2718	&	31.1	&	5.9	&	0.0682 	\\	
$^{ 126 }$Xe	&	0.9195	&	39.4	&	7.8	&	60.59		\\
$^{ 132 }$Ba	&	0.8439	&	42.3	&	7.7	&	61.98		\\
$^{ 138 }$Ce	&	0.6930	&	45.6	&	9.2	&	34.47		\\
$^{ 152 }$Gd	&	0.05570	&	55.9	&	11.6	&	1.121$\times 10^{-2}$	\\	
$^{ 158 }$Dy	&	0.2829	&	59.6	&	12.4	&	3.191		\\
$^{ 164 }$Er	&	0.02506	&	63.4	&	13.3	&	8.309$\times 10^{-3}$	\\	
$^{ 180 }$W	&	0.1433	&	75.8	&	16.0	&	1.4781		\\
$^{ 196 }$Hg	&	0.8206	&	89.9	&	19.5	&	3587 		\\ 
& &  & &  \\ \hline
\end{tabular}
\end{table}


\end{document}
\end